\documentclass[article,prl,balancelastpage,twocolumn,showpacs]{revtex4}
\usepackage{makeidx}
\usepackage{amssymb}
\usepackage{dcolumn}
\usepackage{graphicx}
\usepackage{acronym}

\input{tcilatex}

\begin{document}

\title{Magnetic moment of relativistic fermions}
\author{B. V. Gisin }
\affiliation{IPO, Ha-Tannaim St. 9, Tel-Aviv 69209, Israel. E-mail:
borisg2011@bezeqint.net}
\date{\today }

\begin{abstract}
\noindent In the paper a new class of exact localized solutions of Dirac's
equation in the field of a circularly polarized electromagnetic wave and a
constant magnetic field is presented. These solutions possess unusual
properties and are applicable only to relativistic fermions. The problem of
the magnetic resonance is considered in the framework of the classical
theory of fields. It is shown that interpretation of the magnetic resonance
for relativistic fermions must be changed. Numerical examples of parameters
of the electromagnetic wave, constant magnetic field and the localization
length scale for real measurements are presented.\hfill
\end{abstract}

\pacs{03.65.Ge, 71.70.Di, 13.49.Em\hspace{20cm}}
\maketitle

\section{Introduction}

It is well known that exact solutions of quantum mechanics equations are of
fundamental importance in physics. However, only few such solutions are
known. We present a new class of exact localized solutions of the Dirac
equation in the field of traveling circularly polarized electromagnetic wave
and a constant magnetic field applied along the wave propagation.

It also is well known that the problem of the magnetic moment is of special
interest in physics. Its contemporary value is deduced in the framework of
quantum electrodynamics \cite{Achi} and for electron have been confirmed
with the amazing accuracy in a variety of experiments \cite{quant}. However,
these measurements correspond to not large velocities and not very strong
magnetic fields. The theoretical basis for that is the Pauli equation and
the magnetic resonance interpretation as a flop of spin under action of a
constant and oscillating magnetic field \cite{Lan}. In the Pauli equation
the operator of the magnetic moment is proportional to that of spin. In the
relativistic case this dependence is more complicated.

In the paper we consider this problem and find a exact analytical solution
for this case. Since operators of energy, momentum and spin in a variable
magnetic field do not commute with the Hamiltonian, we must consider average
values of these operators. In fact this consideration is equivalent to that
in the framework of the classical theory of fields \cite{LTP}. We show that
in the relativistic case the magnetic resonance is a flop of the magnetic
moment but not spin.

We consider numerical examples of presented solutions applicable to
experimental testing.

\section{Solutions of Dirac's equation}

Consider Dirac's equation\ 
\begin{equation}
i\hbar \frac{\partial }{\partial t}\Psi =c\mathbf{\alpha }(\mathbf{\mathbf{p}%
}-\frac{e}{c}\mathbf{\mathbf{A}})\Psi +\beta mc^{2}\Psi =0  \label{Dir}
\end{equation}%
in the field of a traveling circularly polarized electromagnetic wave and a
constant magnetic field $H_{z}$ directed along the $z$-axis . Such a field
corresponds to the potential%
\begin{eqnarray}
A_{x} &=&-\frac{1}{2}H_{z}y+\frac{1}{k}H\cos (\Omega t-kz),  \label{Ax} \\
A_{y} &=&\frac{1}{2}H_{z}x+\frac{1}{k}H\sin (\Omega t-kz),  \label{Ay}
\end{eqnarray}%
\ where $k=\varepsilon \Omega /c$ is the propagation constant, $\Omega $ is
the frequency, the sign change of $\Omega $ corresponds to the opposite
polarization, values $\varepsilon =1$ and $\varepsilon =-1$ are used when
the wave propagates along the $z$-axis and opposite direction respectively, $%
c$ is the speed of light, $\alpha _{k},\beta $ are Dirac's matrices, $H$ is
the amplitude of this wave. It is well known that the amplitudes of the
electric and magnetic fields in the plane wave coincide. These fields are an
integral part of the wave and the influence of the electric and magnetic
field cannot be considered separately in the given case.

The transition into a rotating frame is essential for the modulation by a
rotating electromagnetic field since in such a frame stationary solutions
are possible. It is well known that the transformation $\ \Psi ^{\prime
}=\exp (\frac{1}{2}\alpha _{1}\alpha _{2}\Phi )\Psi $ \ describes a spinor
in the frame turned around the $z$-axis through an angle $\Phi .$ If $\Phi
=\Omega t,$\ then \ it describes a spinor in the rotating frame. In
accordance with this, an appropriate coordinate transformation is also
necessary. Below we use the tilde for coordinates in the rotating frame%
\begin{eqnarray}
\tilde{x} &=&r\cos \tilde{\varphi},\text{ \ }\tilde{y}=r\sin \tilde{\varphi}%
,\   \label{xy} \\
\tilde{\varphi} &=&\varphi -\Omega t+kz,\text{ }\tilde{t}=t,\text{ }\tilde{z}%
=z.
\end{eqnarray}%
This transformation is identified as a "Galilean transformation for rotating
frames" because time in the initial and rotating frame is the same.\ 

We use constants $E$ and $p$ as "energy" and "momentum along the $z$-axis"
for stationary states in the rotating frame.\ Once these states have been
found the wave function as well as coordinates are translated back into the
initial (non-rotating) frame.

In the initial frame Eq. (\ref{D}) has exact solutions localized
perpendicularly to the $z$-axis \cite{arx}%
\begin{eqnarray}
\Psi &=&\exp [-i\frac{E}{\hbar }t+i\frac{p}{\hbar }z-\frac{1}{2}\alpha
_{1}\alpha _{2}(\Omega t-kz)+D]\psi ,  \label{psi} \\
D &=&-\frac{1}{2}d(\tilde{x}^{2}+\tilde{y}^{2})+d_{1}\tilde{x}+d_{2}\tilde{y}%
,  \label{D}
\end{eqnarray}%
$\psi $ is a spinor polynomial in $\tilde{x}$ and\ $\tilde{y}.$\ If $\psi $
is a constant spinor then solution (\ref{psi}) describes the "ground state".
Below, as a simplest example, we investigate properties of this state.

Localized solutions (\ref{psi}) exist if the parameter $d$ is positive and
defined by the equality%
\begin{equation}
d^{2}=\frac{e^{2}}{4\hbar ^{2}c^{2}}H_{z}^{2}.  \label{d2}
\end{equation}%
In accordance with Eq. (\ref{d2}), two types of solutions are possible. We
denote them as $\psi _{-}$ \ for $eH_{z}<0$ and $\psi _{+}$ for $eH_{z}>0.$
These normalized spinors of the ground state have the form

\begin{equation}
\psi _{-}=N_{-}\left( 
\begin{array}{c}
h\mathcal{E} \\ 
-\varepsilon (\mathcal{E}+1)(\mathcal{E}-\mathcal{E}_{0}) \\ 
\varepsilon h\mathcal{E} \\ 
-(\mathcal{E}-1)(\mathcal{E}-\mathcal{E}_{0})%
\end{array}%
\right) ,  \label{psim}
\end{equation}

\begin{equation}
\psi _{+}=N_{+}\left( 
\begin{array}{c}
(\mathcal{E}+1)(\mathcal{E}+\mathcal{E}_{0}) \\ 
\varepsilon \mathcal{E}h \\ 
-\varepsilon (\mathcal{E}-1)(\mathcal{E}+\mathcal{E}_{0}) \\ 
-\mathcal{E}h%
\end{array}%
\right) ,  \label{psip}
\end{equation}%
where $N_{\mp }$ is defined by the normalization condition $\int \Psi ^{\ast
}\Psi dxdy=1.$

\begin{equation}
N_{\mp }=\frac{\sqrt{d/2\pi }\exp (-d_{2}^{2}/2d)}{\sqrt{(\mathcal{E}^{2}+1)(%
\mathcal{E}\mp \mathcal{E}_{0})^{2}+h^{2}\mathcal{E}^{2}}},  \label{N}
\end{equation}%
\begin{equation}
d_{1}=\mp id_{2},\text{ \ }d_{2}=\frac{\mathcal{E}_{0}mch}{2\hbar (\mathcal{E%
}\mp \mathcal{E}_{0})}.  \label{dd12}
\end{equation}%
The upper and lower sign before a parameter corresponds to solutions with
negative and positive $eH_{z}$ respectively.

Eigenvalues of "energy in the rotating frame" $E$ are defined with help of a
normalized parameter $\mathcal{E}\equiv -(E-\varepsilon pc)/mc^{2}$. This
parameter obeys the equation%
\begin{equation}
\mathcal{E}^{3}+(\mp \mathcal{E}_{0}+\Lambda _{\mp })\mathcal{E}^{2}-(1\pm 
\mathcal{E}_{0}\Lambda _{\mp }+h^{2})\mathcal{E\pm E}_{0}=0,  \label{E0}
\end{equation}%
where%
\begin{equation}
\text{\ \ \ }\mathcal{E}_{0}\mathcal{=}\frac{2\hbar d}{\Omega m},\text{ \ }%
\Lambda _{\mp }=\frac{2\varepsilon pc\mp \hbar \Omega }{mc^{2}},\text{ \ \ }%
h=\frac{e}{kmc^{2}}H.\text{ \ \ }  \label{vh}
\end{equation}

Obviously, wave functions (\ref{psim}) and (\ref{psip}) cannot be presented
as a small and large two-component spinor. It means that the difference $%
E^{2}-m^{2}c^{2}$ cannot be small. From this an important conclusion
follows: \emph{these solutions correspond only to the relativistic case}.
Another distinguish feature is an algebraic equation of the third order (\ref%
{E0}) for eigenvalues $\mathcal{E}_{k}$ and correspondingly $E_{k}$, $%
k=1,2,3 $.

The wave function $\psi _{\mp }$ depends on four independent normalized
parameters $\mathcal{E}_{0},\Lambda _{\mp },d,h$ which are used below rather
than $\Omega ,p,H_{z},H.$

\section{Average energy and momentum}

It should be emphasized that in the initial frame states are not stationary,
operators of energy and momentum don't commute with the Hamiltonian.
Therefore average values of operators must be used. In the initial frame
average energy $\bar{E}\equiv i\hbar \int \Psi ^{\ast }\frac{\partial }{%
\partial t}\Psi dxdy$ and momentum $\mathbf{\bar{p}}$ $\equiv -i\hbar \int
\Psi ^{\ast }\mathbf{\nabla }\Psi dxdy$ are as follows 
\begin{eqnarray}
\bar{E} &=&E\mp \zeta _{\mp }\pm \hbar \Omega \frac{d_{2}^{2}}{d},\text{ }
\label{Ecv} \\
\bar{p}_{z} &=&p\mp \frac{\varepsilon }{c}\zeta _{\mp }\pm \hbar \Omega 
\frac{\varepsilon d_{2}^{2}}{cd},  \label{paz} \\
\bar{p}_{x} &=&\mp \hbar d_{2}\cos (\Omega t-kz),  \label{pax} \\
\text{ \ }\bar{p}_{y} &=&\pm i\hbar d_{2}\sin (\Omega t-kz),  \label{pay}
\end{eqnarray}%
where 
\begin{equation}
\zeta _{\mp }=\frac{\hbar \Omega }{2}\frac{(\mathcal{E}^{2}+1)(\mathcal{E\mp
E}_{0})^{2}-h^{2}\mathcal{E}^{2}}{(\mathcal{E}^{2}+1)(\mathcal{E\mp E}%
_{0})^{2}+h^{2}\mathcal{E}^{2}}.  \label{N0}
\end{equation}

For states (\ref{psim}) and (\ref{psip}) the uncertainty principle results
in the equality%
\[
\Delta x\Delta p_{x}=\frac{1}{2}\hbar 
\]%
where $\Delta x=\sqrt{1/2d}$ and $\Delta p_{x}=\hbar \sqrt{d/2}$ are
standard deviations of $x$ and momentum $p_{x}.$ All that is valid also for
deviations of $y$ and momentum $p_{y}$.

\section{The magnetic moment}

It is well known that the magnetic moment plays a principal role in the
interaction of fermions with a magnetic/electromagnetic field. For
relativistic fermions described here the magnetic moment is determined as
the first derivative of Dirac's Lagrangian with respect to the magnetic
field.\ Since solutions (\ref{psi}) are localized perpendicularly to the $z$%
-axis, it is appropriate to consider only \ the $z$-component of the
magnetic moment. This component is 
\begin{equation}
\mu _{z}=\frac{e}{2}\int (y\Psi ^{\ast }\alpha _{1}\Psi -x\Psi ^{\ast
}\alpha _{2}\Psi )dxdy.  \label{mag}
\end{equation}%
This definition coincides with that in classical theory of fields \cite{LTP}%
. Below we study this quantity contrary to the classical consideration where
a constant in the the Pauli equation is used instead $\mu _{z}$ .

In non-relativistic case a rotating electromagnetic field gives rise
electron pairs with spins rotating in opposite direction \cite{arx}. The
wave function of electron is formed from two wave functions corresponding
two stationary states in the rotating frame with a certain difference
between energy levels. In the given case the wave function also has to be
constructed from two functions%
\begin{equation}
\Psi =C_{1}\Psi _{1}+C_{2}\Psi _{2},  \label{psi1}
\end{equation}%
where $C_{k}$ is the normalization coefficient%
\[
|C_{1}|^{2}+|C_{2}|^{2}=1. 
\]

The wave function (\ref{psi}) depends on four parameters $\mathcal{E}%
_{0},\Lambda _{\mp },d,h$ and two roots of Eq. (\ref{E0}). For the study of
magnetic moment properties it is convenient to introduce two new parameters%
\begin{equation}
\Pi =\mathcal{E}_{1}\mathcal{E}_{2},\text{ \ }\eta =\mathcal{E}_{1}+\mathcal{%
E}_{2}.  \label{np}
\end{equation}%
Then from Eq. (\ref{E0}) one follows%
\begin{equation}
h^{2}=-\frac{\Pi +1}{\Pi }(1\mp \mathcal{E}_{0}\eta +\mathcal{E}_{0}^{2}),%
\text{ \ }\Lambda _{\mp }=\pm \mathcal{E}_{0}-\eta \mp \frac{1}{\Pi }%
\mathcal{E}_{0}
\end{equation}%
Four new variable parameters $\mathcal{E}_{0},\Pi ,\eta ,d$ can be
considered as independent. With these parameters there is not need to refer
to Eq. (\ref{E0}).

Substitute (\ref{psi1}) in Eq. (\ref{mag}). The integration in (\ref{mag})
results in a exact expression. This expression consists of a
time-independent and oscillating part. The constant part may be reduced to
zero if $\Pi <0,$ i.e., if $\mathcal{E}_{1}$ and $\mathcal{E}_{2}$ have
opposite signs. The variable part oscillates at the frequency%
\begin{equation}
\frac{E_{1}-E_{2}}{\hbar }.  \label{frec}
\end{equation}%
The dependence on $d$ may be extracted from the variable part:%
\begin{equation}
\frac{1}{d}\exp [-\frac{(d_{2}^{\prime }-d_{2}^{\prime \prime })^{2}}{2d}],
\label{fd}
\end{equation}%
where $d_{2}^{\prime }$ and $d_{2}^{\prime \prime }$ are independent on $d$ (%
\ref{dd12}) and correspond to states with $\mathcal{E}_{1}$ and $\mathcal{E}%
_{2}$ respectively.

\section{The magnetic resonance}

It is well known that in non-relativistic case the magnetic resonance occurs
at a value of the longitudinal magnetic field corresponding to the spin
oscillation amplitude maximum. At this value the amplitude becomes constant.

Such a maximization in relativistic case corresponds to an extremum of the
function (\ref{fd}) in respect to $d$. At the extremum point 
\begin{equation}
2d=(d_{2}^{\prime }-d_{2}^{\prime \prime })^{2}.  \label{dm}
\end{equation}%
With help of this equality the parameter $d$ may be excluded from
consideration. However, in contrast to the classical case, the amplitude of
the magnetic moment oscillation remains dependent on thee parameters $\Pi ,%
\mathcal{E}_{0},\eta $. The change of these parameters allows to vary the
magnetic moment over a wide range. The exact expression for the oscillation
amplitude  $A_{\mp }$ at the standard classical condition $%
|C_{1}|^{2}=|C_{2}|=\frac{1}{2}$ takes the form%
\begin{eqnarray}
A_{-} &=&\mu _{B}\frac{4R_{-}}{\mathcal{E}_{0}\exp (1)}[-2\Pi +2\mathcal{E}%
_{0}\eta -2\mathcal{E}_{0}^{2})],  \label{Am} \\
A_{+} &=&\mu _{B}\frac{4R_{+}}{\mathcal{E}_{0}\exp (1)}[2\Pi -2\eta \mathcal{%
E}_{0}-\eta ^{2}-2\mathcal{E}_{0}{}^{2}],  \label{Ap} \\
R_{\mp } &=&\sqrt{\frac{-\Pi ^{3}(\Pi \mp \mathcal{E}_{0}\eta +\mathcal{E}%
_{0}^{2})}{(\Pi ^{3}\pm \Pi \eta \mathcal{E}_{0}+\mathcal{E}_{0}^{2})(\eta
^{2}-4\Pi )^{3}}}.
\end{eqnarray}%
$\mu _{B}$ is the Bohr magneton.

Other pair of wave functions may be considered analogously. But the
maximization is possible only for one pair of wave functions.

It is easy to show that the magnetic moment oscillation frequency (\ref{frec}%
), in contrast to the classical case, depends on the fermion mass (and $%
\mathcal{E}_{0},\Pi ,\eta $ or non-normalized parameters $\Omega ,p,h$).
Therefore, measurements of this frequency probably may be turned into the
precision measurement of fermion masses, provided an experimental technique
will be developed for the estimate of the number of oscillations and fermion
time-of-flight. Particular cases may be calculated with help of exact
expressions for $A_{\mp }$.

An consideration of the average spin $s_{3}=-\frac{i}{2}\hbar \int \Psi
^{\ast }\alpha _{1}\alpha _{2}\Psi dxdy$ shows that $s_{3}$ also consists of
a constant and oscillating part. The time-dependent part oscillates with the
same frequency (\ref{frec}), however, the amplitude of oscillations has no
an extremum by varying $H_{z}$ and monotonically decreases to zero at the
change of $H_{z}$ from infinity to zero. From this the second important
conclusion follows: \emph{in the relativistic case the magnetic resonance is
the flop of the magnetic moment but not spin.}

\section{Some examples{}}

In this Section we consider particular values of \ $H_{z},\Omega ,H$ for
which a real testing of presented solutions is possible. For definiteness we
consider examples for electron.

We start from such a characteristic feature as the localization length scale 
$l.$ This scale perpendicular to the $z$-axis is defined by the parameter $d$
(\ref{d2}) 
\begin{equation}
l\approx 2\sqrt{\frac{1}{d}}=2\sqrt{|\frac{2\hbar c}{eH_{z}}|}.  \label{l}
\end{equation}

The frequency $\omega =\Omega /2\pi $ may be found from particular values of 
$\mathcal{E}_{0}$ (\ref{vh})$.$ The first example illustrates the $l$- and $%
\omega $- dependence of $H_{z}$ from relatively weak to strong values$.$ The
magnetic field of the order of $40T$ \ is used as an upper limit of the
man-made magnetic field producing the scale lower limit. In this example $%
\mathcal{E}_{0}=1.$ 
\begin{eqnarray*}
&&%
\begin{array}{c}
H_{z}(G) \\ 
l(cm) \\ 
\omega (Hz)%
\end{array}%
\left\vert 
\begin{array}{c}
1 \\ 
7.26\cdot 10^{-4} \\ 
2.80\cdot 10^{6}%
\end{array}%
\right\vert 
\begin{array}{c}
10^{3} \\ 
2.29\cdot 10^{-5} \\ 
2.80\cdot 10^{9}%
\end{array}%
\left\vert 
\begin{array}{c}
4\cdot 10^{5} \\ 
1.15\cdot 10^{-6} \\ 
1.12\cdot 10^{12}%
\end{array}%
\right\vert \\
&&\text{\ \ \ \ \ \ \ \ \ \ \ \ \ \ \ \ \ \ \ \ \ \ \ \ \ \ \ \ \ Table 1}
\end{eqnarray*}%
These dependences are not connected with the magnetic resonance.

The magnetic resonance places the constraint on $d$ (\ref{dm}). The second
example illustrates $\omega $- and $h$- dependence of $H_{z}$ in the
condition of the magnetic resonance and at the condition 
\begin{equation}
|\mathcal{E}_{0}|\ll 1.  \label{con}
\end{equation}%
This condition enables to simplify calculations. Moreover, the inequality $%
|\hbar \Omega |/mc^{2}\ll 1$ is always holds. Therefore, for relativistic
fermions $\Lambda _{\mp }=2\varepsilon p/mc$. In this example $|\mathcal{E}%
_{0}|=0.4117\cdot 10^{-2}$ is chosen so that $\omega $ corresponds to the
frequency of a powerful\ Nd:YAG laser with the wavelength of $1.064$ micron.%
\begin{eqnarray*}
&&%
\begin{array}{c}
H_{z}(G) \\ 
\omega (Hz) \\ 
hP%
\end{array}%
\left\vert 
\begin{array}{c}
1 \\ 
7.04\cdot 10^{7} \\ 
3.66\cdot 10^{-5}%
\end{array}%
\right\vert 
\begin{array}{c}
10^{3} \\ 
7.04\cdot 10^{10} \\ 
1.17\cdot 10^{-3}%
\end{array}%
\left\vert 
\begin{array}{c}
4\cdot 10^{5} \\ 
2.82\cdot 10^{14} \\ 
2.31\cdot 10^{-2}%
\end{array}%
\right\vert \\
&&\text{ \ \ \ \ \ \ \ \ \ \ \ \ \ \ \ \ \ \ \ \ \ \ \ \ \ \ \ \ \ Table 2}
\end{eqnarray*}%
where $P=\sqrt{p^{2}/m^{2}c^{2}+1}$. A corresponding value of $H$ may be
found from the relation $H=\varepsilon hH_{z}/\mathcal{E}_{0}.$

The laser radiation may be transformed by means of optics in a beam with the
cross-section of the order of few microns. This cross-section may be much
more than $l$. The energy density by this transformation increases by a
factor of the order of $10^{8}$ times. Such data may be used for flows of
relativistic fermions of a very small cross-section driven by a powerful
laser beam.

Above examples far not exhaust all possibilities for experiments.\ \ \
\smallskip\ \ \ \ \ \ \ \ \ \ \ \ \ \ \ \ \ \ \ \ \ \ \ \ \ \ \ \ \ \ \ \ \
\ \ \ \ \ \ \ \ \ \ \ \ \ \ \ \ \ \ \ \ 

\section{Conclusion}

We have considered the new class of exact solutions of the Dirac equation
corresponding to relativistic fermions. Practical numerical examples
presented in the paper can describe, in particular, flows of relativistic
fermions of a very small cross section driven by a powerful laser beam in
accelerators and colliders. We have found the exact analytical solution for
the problem of the magnetic resonance and shown that in relativistic case
the magnetic resonance is a flop of the magnetic moment but not spin.
Perhaps such a resonance may be turned into precision measurements of
fermion masses. Considered solutions may be an useful and powerful
instrument as for high energy physics as for astrophysics, in particular,
for the description of processes near neutron stars and magnetars.

\end{document}